\documentclass{pnasthree}

\usepackage{amssymb,amsfonts,amsmath}
\usepackage{color}
\usepackage{graphicx}
\usepackage{epsfig}
\usepackage{booktabs}
\usepackage[normalem]{ulem}

\newcommand{\req}[1]{Eq.~(\ref{#1})}
\newcommand{\avg}[1]{{\langle #1\rangle}}

\def\cL{{\cal L}}
\def\cH{{\cal H}}

\def\cZ{{\cal Z}}

\newcommand{\vS}{{\vec{S}}}
\newcommand{\vSi}{{\vec{S}_{i}}}

\newcommand{\scom}{{a}}
\newcommand{\kone}{{k_1}}
\newcommand{\ktwo}{{k_2}}
\newcommand{\kL}{{k_l}}

\newcommand{\lin}{{\lambda_i^\nu}}
\newcommand{\ljn}{{\lambda_j^\nu}}

\newcommand{\ljns}{{\lambda_j^{\nu*}}}
\newcommand{\Ln}{{\Lambda_\nu}}
\newcommand{\Lm}{{\Lambda_\mu}}

\newcommand{\ajin}{{a_{j\to i}^\nu}}

\newcommand{\aljn}{{a_{l\to j}^\nu}}

\newcommand{\aljm}{{a_{l\to j}^\mu}}

\newcommand{\bjin}{{b_{j\to i}^\nu}}

\newcommand{\bljn}{{b_{l\to j}^\nu}}
\newcommand{\brjn}{{b_{r\to j}^\nu}}

\newcommand{\brjm}{{b_{r\to j}^\mu}}

\newcommand{\Ljn}{{\Lambda_j^\nu}}

\newcommand{\flow}{{I}}
\newcommand{\flowi}{{I_i}}
\newcommand{\invS}{{\int_\odot d \vS}}
\newcommand{\invSi}{{\int_\odot d \vSi}}

\newcommand{\Ep}{{E_{\rm P}}}
\newcommand{\Ed}{{E_{\rm D}}}
\newcommand{\Emc}{{E_{\rm MC}}}
\newcommand{\Lp}{{L_{\rm P}}}
\newcommand{\Ld}{{L_{\rm D}}}
\newcommand{\Lmc}{{L_{\rm MC}}}

\newcommand{\cut}[1]{{}}

\contributor{Submitted to Proceedings
of the National Academy of Sciences of the United States of America}
\url{www.pnas.org/cgi/doi/10.1073/pnas.0709640104}
\copyrightyear{2008}
\issuedate{Issue Date}
\volume{Volume}
\issuenumber{Issue Number}

\begin{document}



\title{From the Physics of Interacting Polymers to Optimizing Routes on the London Underground}





\author{Chi Ho Yeung\affil{1}{The Nonlinearity and Complexity Research Group, Aston University, Birmingham B4 7ET, United Kingdom},
David Saad\affil{1}{}, \and
K. Y. Michael Wong\affil{2}{Department of Physics, The Hong Kong University of Science and Technology, Hong Kong}
}

\contributor{Submitted to Proceedings of the National Academy of Sciences
of the United States of America}

\maketitle

\begin{article}

\begin{abstract}
Optimizing paths on networks is crucial for many applications, from subway traffic to Internet communication. As global path optimization that takes account of all path-choices simultaneously is computationally hard, most existing routing algorithms optimize paths individually, thus providing sub-optimal solutions. We employ the physics of interacting polymers and disordered systems to analyze macroscopic properties of generic path-optimization problems and derive a simple, principled, generic and distributed routing algorithm capable of considering simultaneously all individual path choices. We demonstrate the efficacy of the new algorithm by applying it to: (i) random graphs resembling Internet overlay networks; (ii) travel on the London underground network based on Oyster-card data; and (iii) the global airport network. Analytically derived macroscopic properties give rise to insightful new routing phenomena, including phase transitions and scaling laws, which facilitate better understanding of the appropriate operational regimes and their limitations that are difficult to obtain otherwise.
 \end{abstract}






\section{Introduction}

Path optimization affects many of our daily activities. While much attention has been dedicated to routing algorithms for Internet applications such as instant messengers and peer-to-peer systems~\cite{huitema95,moy98}, many other essential routing applications have attracted less attention; from water distribution networks~\cite{vasan10}, sensor networks~\cite{rangwala06}, military convoy movements~\cite{chardaire05} to journey planners~\cite{beckmann56,wardrop52}. In many applications, enormous costs are incurred due to traffic congestion or non-essential and redundant capacity. Due to the computational costs involved, most existing routing algorithms are static and based on selfish decisions, with non-adaptive routing tables indicating the shortest path to destinations regardless of local traffic~\cite{bellman58,dijkstra59}. Dynamic routing protocols do exist, but they are either heuristic, probabilistic or insensitive to other individual path decisions which dynamically constitutes the traffic~\cite{xie04,zhu06}. A more global approach that takes into account all individual path decisions is crucial for efficient use of over-stretched infrastructure. For instance, one may suppress congestion by minimizing overlaps with other routes, or decrease the number of active nodes by consolidating paths to reduce infrastructure demands or energy consumption. The latter is particularly important in the context of the Internet as it can consume up to 4$\%$ of the electricity generated~\cite{baliga07}. Future applications include individualized routing and optimal resource management of pre-booked air and road traffic.

The difficulty in deriving a globally-optimal algorithm, in contrast to greedy local ones, lies in the simultaneous assignment of multiple interacting paths to minimize a global cost, as the optimal path between any particular source-destination pair depends on all other paths choices. Such interaction is highly non-local, as paths between different source-destination pairs may partially overlap. Existing algorithms either ignore these interactions~\cite{bellman58,dijkstra59}, or use heuristics to approximate them~\cite{xie04,zhu06}; both approaches result in sub-optimal solutions. A substantial effort has been devoted to the development of highly efficient routing methods, for instance multi-commodity flow algorithms~\cite{shahrokhi90,leighton88,awerbuch93,garg98,awerbuch07,barnhart00,castro96}. However, most methods are based on weighted linear objective functions and real variables and aim specifically at satisfying capacity constraints; they have limited flexibility in addressing the variety of non-linear cost functions one may want to optimize in different scenarios, especially concave costs and integer variables. A more detailed discussion is provided in Section S4 of the \emph{SI Appendix}.

Here we employ statistical physics-based methods used in the study of interacting polymers~\cite{daoud75} and spin glasses~\cite{mezard87,nishimori01} to obtain both a macroscopic description of the routing problem and microscopic solutions for given instances; the latter leads to a simple, generic and distributed routing optimization algorithm. The algorithm resembles message passing techniques that have been developed independently in a number of disciplines~\cite{mezard87,Pearl,GallagerBook} and have been successfully applied to a variety of problems from prototyping~\cite{Frey10} to solving hard computational problems~\cite{mezard02} and control of complex systems~\cite{Barabasi11}. Here we demonstrate the potential and efficacy of our routing algorithm by applying it to random networks, individualized routing on the London subway network and the global airport network. Together with other benchmark tests described in the \emph{SI Appendix}, we demonstrate that our algorithm achieves better optimization compared to existing heuristics and state-of-the-art approximation algorithms in various routing scenarios; moreover, it is distributed, principled and does not require fine-tuning of free parameters.

In addition to the significant algorithmic advances, several macroscopic phenomena including a phase transition, scaling rules as a function of network size and non-monotonic growth in mean path length as a function of traffic volume are revealed; these cannot be obtained by numerical studies and provide new insights and understanding of optimal routing on sparse networks.

\section{Model}

Consider a system of $M$ polymers interacting on a network of $N$ nodes. Each node $i\!=\!1,\ldots, N$ is connected to $k_i$ neighbors denoted by the set $\cL_i$ and the connectivity matrix $A_{ij}=A_{ji}=1$ when $i$ and $j$ are connected and zero otherwise. Each polymer $\nu\!=\!1,\ldots, M$ has two fixed ends and occupies a path described by a self-avoiding walk on the network, i.e. consecutive segments occupy topological neighbors and each polymer $\nu$ goes through a node at most once. We denote the variable $\sigma_i^\nu=1$ when polymer $\nu$ occupies node $i$ and $\sigma_i^\nu=0$ otherwise, and the number of polymers occupying $i$ as $\flow_i=\sum_\nu\sigma_i^\nu$. To penalize or encourage polymer overlap, we define the Hamiltonian $\cH$ to be a non-linear function of the normalized flow $\lambda_i=\flow_i/M$, namely
\begin{align}
	\cH=M\sum_i\phi(\lambda_i)~. \label{eq_Hamiltonian}
\end{align}
The analytic solution and derived algorithm are generic for any $\phi$. While the current framework focuses on undirected polymers and costs which incur at vertices, it is clear that in some applications costs incur at the edges and edges may be directed and weighted. Our framework, derivation and algorithms accommodate costs on edges (using a factor graph representation) as well as directed and weighted polymers, making them suitable for most routing scenarios. The derivation and corresponding algorithms are given in \emph{SI Appendix} Sec.~S3. We would like to point out that the algorithm presented below already accommodates directed traffic.

This model is equivalent to a setting of $M$ source-destination pairs, which we term \emph{communications}, each of which occupies a path on a network with $N$ nodes. The variable $\lambda_i$ is thus the normalized traffic on node $i$ and $\cH$ the corresponding cost function. In the physical framework and the zero-temperature limit, we minimize $\cH$ to obtain the ground state of the system or the optimal path configuration of the corresponding routing system. Some simple forms of $\cH$ are already meaningful, for instance $\phi(x)=x^\gamma$, where the cases with $\gamma>1$ penalize overlaps to suppress congestion while $\gamma<1$ encourages overlaps to aggregate traffic~\cite{bohn07, banavar00, shao07}. The case of $\gamma=1$ reduces to  $\cH\propto\sum_\nu(\sum_i \sigma_i^\nu)$ whose ground state corresponds to shortest-path routing.

\section{Methods}
\subsection{Theoretical Approach}

The main obstacle in accounting for the interaction between paths is in keeping track of the cost at local nodes or edges while maintaining path-integrity between the two end points and avoiding redundant loops. Therefore, in addition to the cost at the various nodes, given by \req{eq_Hamiltonian}, we introduced a technique used in polymer physics~\cite{daoud75}, in the study of self-avoiding walks~\cite{batchelor89, stilick96}, to enforce the appropriate path constraints.

The method is based on representing each node as an $n$-component vector $\vS$ of length $\sqrt{n}$.
Denoting the angular integration over $\vS$ as $\invS$, it has been shown~\cite{daoud75} that all positive moments of $S_\scom$ vanish in the limit $n\to 0$ except the second moment $\frac{1}{C_n}\invS S_\scom^2=1$ for any component $\scom$ in $\vS$, where $C_n=\invS$ is a normalization constant. It then implies that when $n\to 0$ all nonvanishing terms that contribute in
\begin{align}
\label{eq_pathCount}
\prod_{i=1}^N\left(\frac{1}{C_n}\invSi\right)
S_{x,\scom}S_{y,\scom}\prod_{(kl)}\left(1+A_{kl}\vec{S}_k\cdot\vec{S}_l\right)
\end{align}
are of the form $A_{x \kone}A_{\kone \ktwo}\cdots A_{\kL y}S^2_{x,\scom}S^2_{\kone,\scom}S^2_{\ktwo,\scom}\cdots S^2_{\kL,\scom}S^2_{y,\scom}$, where $k_i$ represents the $i$-th node index of the corresponding path/polymer segment; these sequences represent self-avoiding paths over nodes $(x, \kone, \ktwo, \dots, \kL, y)$, joining the end nodes $x$ and $y$~\cite{daoud75}. Each node that is part of these paths incurs a cost as in \req{eq_Hamiltonian}; a sum over all possible paths of all communications provides the partition function $\cZ$, as detailed in \emph{SI Appendix} Sec.~S1.1. To obtain typical macroscopic properties one needs to average $\cZ$ over topologies (given a degree distribution) and node-pair choices, termed \emph{quenched disorders} in statistical physics. This requires the use of the replica or cavity methods of spin glass theory \cite{mezard87,nishimori01}, as presented in \emph{SI Appendix} Sec.~S1.

The aim of the analysis is two-fold: (1) At the macroscopic level, we derive the stable traffic distribution $P(\flow)$ in the limit of very large systems to obtain the average cost (energy) $\avg{E}=\avg{\phi(\flow/M)}$, the average path length, given by the total occupancy divided by $M$, i.e. $\avg{L}=\frac{N}{M}\avg{\flow}$, and the average fraction of idle nodes given by $f_{\rm idle}=\avg{\delta(\flow)}$, detailed in  \emph{SI Appendix} Sec.~S1.4. Angled brackets denote an average over $P(\flow)$, which includes averages over all variable states for a given network and over choices of network and end-point instances. (2) At the microscopic level, the cavity based analysis~\cite{mezard09} translates to an algorithm which optimizes path configuration in a principled, distributed and computationally efficient manner.

\subsection{Optimization algorithm}

The analytical solutions for infinite systems translate into an optimization algorithm valid for finite systems, as detailed in \emph{SI Appendix} Sec.~S2. The derived algorithm is based on sending a couple of messages $\ajin$ and $\bjin$ at the zero temperature limit, from node $j$ to node $i$ for each index $\nu$; these characterize the energy contributions of communication $\nu$ at edge $j\rightarrow i$, originated from the source and destination directions, respectively. The messages take the form:
\begin{align}
\label{eq_a}
&\ajin =
\nonumber\\
&\begin{cases}
\displaystyle
\min_{l\in \cL_j\backslash \{i\}}\left[\aljn\right]
-\min\left[-\phi'(\ljns), \min_{\substack{l,r\in \cL_j\backslash\{i\}\\l\neq r}}\left[\aljn+\brjn\right]\right],
\\
& \hspace{-1.3cm}\Ljn=0
\\
\displaystyle
-\min_{l\in \cL_j\backslash \{i\}}\left[\bljn\right],
& \hspace{-1.3cm}\Ljn=1
\\
\displaystyle
\infty,
& \hspace{-1.5cm}\Ljn=-1
\end{cases}
\\
\label{eq_b}
&\bjin =
\nonumber\\
&\begin{cases}
\displaystyle
\min_{l\in \cL_j\backslash \{i\}}\left[\bljn\right]
-\min\left[-\phi'(\ljns), \min_{\substack{l,r\in \cL_j\backslash\{i\}\\l\neq r}}\left[\aljn+\brjn\right]\right],
\\
&  \hspace{-1.3cm}\Ljn=0
\\
\displaystyle
\infty,
&  \hspace{-1.3cm}\Ljn =1
\\
\displaystyle
0,
&  \hspace{-1.5cm}\Ljn=-1
\end{cases}
\end{align}
where $\Ljn=+1, -1$ for source and destination, respectively, and is zero otherwise; the general cost function  $\phi$ and the set of nodes in the neighborhood of node $j$ is denoted as $\cL_j$. The value of $\ljns$ is given by the solution of $\ljn$ in
\begin{align}
\label{eq_gx}
&\ljn =\frac{1}{M} +\frac{1}{M}\sum_{\mu\neq\nu}\Bigg\{|\Lm|
\nonumber\\
&\qquad\quad\left.+(1-|\Lm|)\Theta\left(-\phi'(\ljn)
-\min_{\substack{l,r\in\cL_j \\l\neq r}}\left[\aljm+\brjm\right]
\right)\right\},
\end{align}
The step function $\Theta(x)$ takes values $\Theta(x)=0,0.5,1$ for $x<0,~x=0$ and $x>0$, respectively.  Solutions of~\req{eq_gx} are obtained by setting $\lin=\flow/M$ and a test integer $\flow$ starting from $\flow=0$ until a self-consistent $\ljn$ is found. Finally, after the set of messages in Eqs.~(\ref{eq_a}) and (\ref{eq_b}) converges to non-fluctuating values, the optimal configuration of path $\nu$ on each node $j$ is given by
\begin{align}
\sigma_j^\nu=|\Ln|+(1-|\Ln|)\Theta\left(-\phi'\left(\ljns\right)
-\min_{\substack{l,r\in\cL_j\\l\neq r}}\left[\aljn+\brjn\right]
\right),
\end{align}
where $\ljns$ is the solution of \req{eq_gx} after convergence, and $\sigma_j^\nu=1$ if the communication $\nu$ passes through node $j$ and zero otherwise. The generalized algorithms which accommodate weighted and directed communications, generic costs on nodes and edges, as well as separate costs defined on directed edges are given in \emph{SI Appendix} Sec.~S3. The computational complexities of these algorithms are discussed in \emph{SI Appendix} Sec.~S2.2.

In some instances the iterative equations fail to converge, this suggests that solution space in the infinite system case is fragmented and non-ergodic; this corresponds to\emph{replica symmetry breaking}~(RSB)~\cite{mezard87,nishimori01},a complicated energy landscape with numerous local minima that typically hinder algorithmic convergence (details in \emph{SI Appendix} Sec.~S5). This is typical in the case of hard computational problems. Convergence is improved by assigning a random bias $\epsilon_i$ to each node~\cite{yeung12}, akin to an external field, guiding the system to one of the local minima. These biases can be easily incorporated in the present formulism by replacing $\phi(x)$ with $\phi_i(x)$ for each node $i$ such that $\phi_i(x)=\phi(x)+x\epsilon_i$. In cases where a large number of source-destination pairs are identical, we further replace $\epsilon_i$ by $\epsilon^\nu_{i}$ for each communication $\nu$ to break the degeneracy brought about by \req{eq_gx}. Details can be found in \emph{SI Appendix} Sec.~S2.1. 
\cut{
The obtained local minima typically provide close-to-optimal solutions.}

\section{Results}
\subsection{Microscopic Solution - Finding Best Paths}
\label{sec_micro}

Employing the suggested algorithm, we can optimize path choices using the cost $\cH\propto\sum_i\flowi^\gamma$. We illustrate the characteristic results obtained by applying the algorithm using two costs, with $\gamma=2$ (convex, $\gamma>1$) and $\gamma=0.5$ (concave, $\gamma<1$), to a system of 10 source-destination pairs communicating on a random regular graph with $N=50$ and $k=3$ as shown in~Fig. 1.

\begin{figure}
\centerline{\epsfig{figure=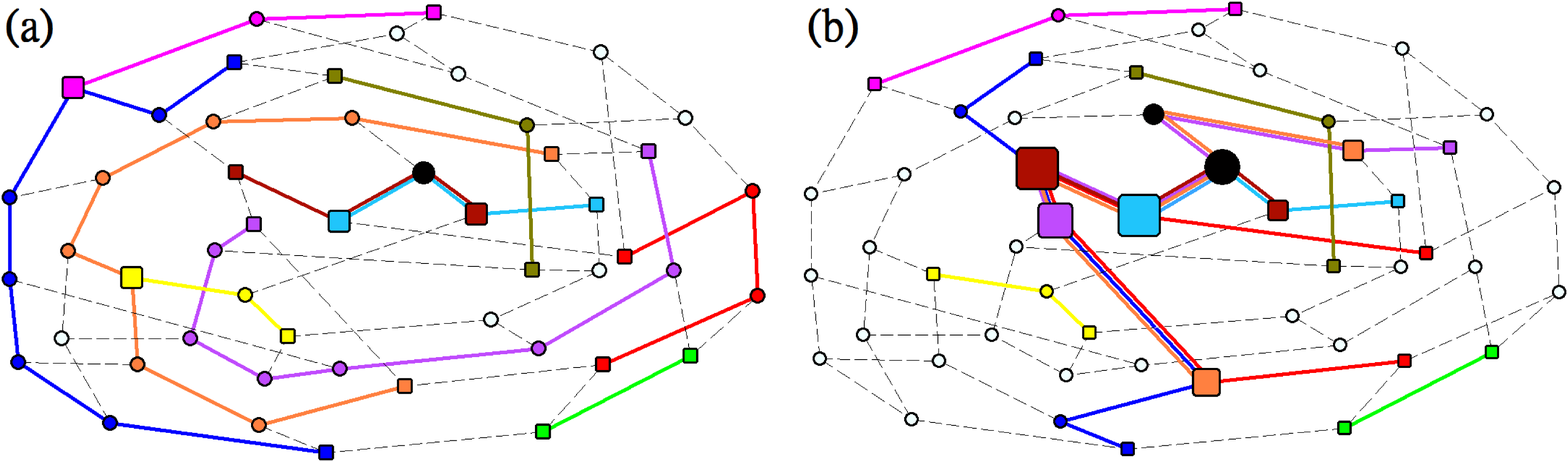, width=\linewidth}}
\caption{
Optimized path configurations on a regular random network. The network comprises 50 nodes (each with $k=3$) and 10 source-destination pairs. The corresponding costs are (a) $\cH\propto\sum_i \flowi^2$, and (b) $\cH\propto\sum_i \flowi^{0.5}$. The path of each communication is illustrated by nodes and edges of a specific color, while black nodes are shared by more than one path. The size of a node is proportional to the amount of traffic through it, and square nodes represent source or destination of each communication. }
\label{fig_ex}
\end{figure}

Figure~1(a) demonstrates how a cost with $\gamma>1$ penalizes congestion: the blue, orange and violet communications are routed via non-shortest paths to avoid overlap, especially in the central congested part of the network. This holds when traffic is heavy and one aims to distribute it uniformly. In contrast to the reduced-congestion solutions, Fig.~1(b) shows solutions obtained for $\cH\propto\sum_i\flowi^{0.5}$, aimed at concentrating traffic. More specifically, the blue, orange and violet communications are all routed via the central congested part of the network that mainly consists of source and destination nodes, making best use of these nodes as relays and leaves many of the other nodes idle. In the case of the Internet or transportation networks, idle nodes can be switched off to save resources.

To demonstrate the efficacy of the algorithm for more realistic systems we examine the performance of the algorithm on the London subway network based on real passenger source-destination data obtained by the Oyster card system~\cite{subwayData}.
We report results for vertex costs only, but similar pictures have been obtained for edge costs and directed traffic.
Figure~3(a) shows how congestion is reduced by the algorithm when $\gamma=2$ is used and traffic is fairly uniform even in the central region (see inset), at the cost of longer individual routes for global optimization.
Table~1 shows that the cost $E=\sum_i\flow_i^2$ obtained by our algorithm is 20.5$\%$ smaller than that of the shortest path configuration obtained by the commonly used Dijkstra algorithm~\cite{dijkstra59}, with only a slight increase in average path length by 5.8$\%$. Practically, traffic optimization of this type may be achieved through differential pricing, or by auxiliary information provided either individually or globally.
On the other hand, when $\gamma=0.5$ is used, paths for the same passenger set are consolidated at major routes and stations as shown in Fig.~3(b). While the size of some of the nodes increases, other branches such as the ones passing through  ``Holborn" and ``Great Portland Street" (see inset) are all but idle. This scenario may be relevant at times when the service is reduced for some reason, for instance a strike or at late evening; service on the shared branches can remain active while the frequency of other less-loaded services decreases.

To better compare the solutions obtained in the two scenarios, we plotted the corresponding traffic at individual stations for the London underground data set in descending order (for  $\gamma=0.5$) as shown in the inset of~Fig.~2. The optimized states of $\gamma=2$ show less traffic for overloaded stations and higher traffic for less-loaded ones; for instance, ``Green Park".

\begin{figure}
\centerline{\epsfig{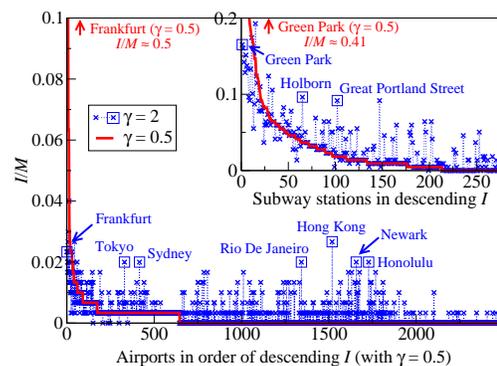}}
\caption{
Optimized traffic at individual airports and London subway stations (inset). The airports and stations are plotted in descending order of traffic in the optimized state of $\cH\propto\sum_i\flowi^{0.5}$ (red lines). Symbols ($\times$) in blue correspond to the optimized traffic with $\cH\propto\sum_i\flowi^{2}$. Squared symbols refer to airports and stations mentioned in the text that are much higher than the red lines. The optimized airport traffic is obtained from the single instance shown in Fig.~5 and the optimized subway traffic is obtained by averaging over the 30 passenger sets as in Table~1.}
\label{fig_traffic}
\end{figure}

\begin{table*}
\begin{center}
\begin{tabular}{@{}lccccr}
\toprule
&\multicolumn{2}{c}{$\gamma=2$} & \phantom{ab} & \multicolumn{2}{c}{$\gamma=0.5$} \\
\cmidrule{2-3}\cmidrule{5-6}
& $\frac{ \Ep-\Ed }{\Ed}$ &  $\frac{ \Lp-\Ld }{\Ld}$ && $\frac{ \Ep-\Ed }{\Ed}$ &  \multicolumn{1}{c}{$\frac{ \Lp-\Ld }{\Ld}$} \\
\midrule
London subway network & $-20.5\pm 0.5\%$ & $+5.8\pm 0.1\%$ && $-4.0\pm 0.1\%$ & $+5.8\pm 0.3\%$ \\
Global airport network & $-56.0\pm 2.0\%$ & $+6.2\pm 0.2\%$ && $-9.5\pm 0.2\%$ & $+8.6\pm 1.2\%$ \\
\midrule
&&&&
\\
& $\frac{ \Ep-\Emc(\alpha^*) }{\Emc(\alpha^*)}$ &  $\frac{ \Lp-\Lmc(\alpha^*) }{\Lmc(\alpha^*)}$ && $\frac{ \Ep-\Emc(\alpha^*) }{\Emc(\alpha^*)}$ &  \multicolumn{1}{c}{$\frac{ \Lp-\Lmc(\alpha^*) }{\Lmc(\alpha^*)}$} \\
\midrule
London subway network & $-0.70\pm 0.04\%$ & $+0.72\pm 0.10\%$ && \multicolumn{2}{c}{No existing algorithm} \\
Global airport network & $-3.09\pm 0.59\%$ & $+0.90\pm 0.64\%$ &&\multicolumn{2}{c}{for comparsion} \\
\bottomrule
\end{tabular}
\end{center}
\textbf{Table 1.} {\small A comparison of average cost $E = \sum_i\flow_i^\gamma$ and path length $L=\frac{1}{M}\sum_i\flow_i$ obtained by our algorithm (P), the Dijkstra algorithm (D) and the modified min-cap congestion aware algorithm (MC)~\cite{shahrokhi90} at individual optimal $\alpha^*$ for each instance. Results are averaged over sets of source-destination pairs recorded in each 1 minute interval between 8:30 am -- 9:00 am on one Wednesday in November 2009 for the London subway network, and 5 sets of 300 randomly drawn source-destination pairs for the global airport network. The values after the $\pm$ signs indicate to the corresponding standard error.}
\label{tab_energyDiff}
\end{table*}

\begin{figure*}
\begin{center}
\centerline{\epsfig{figure=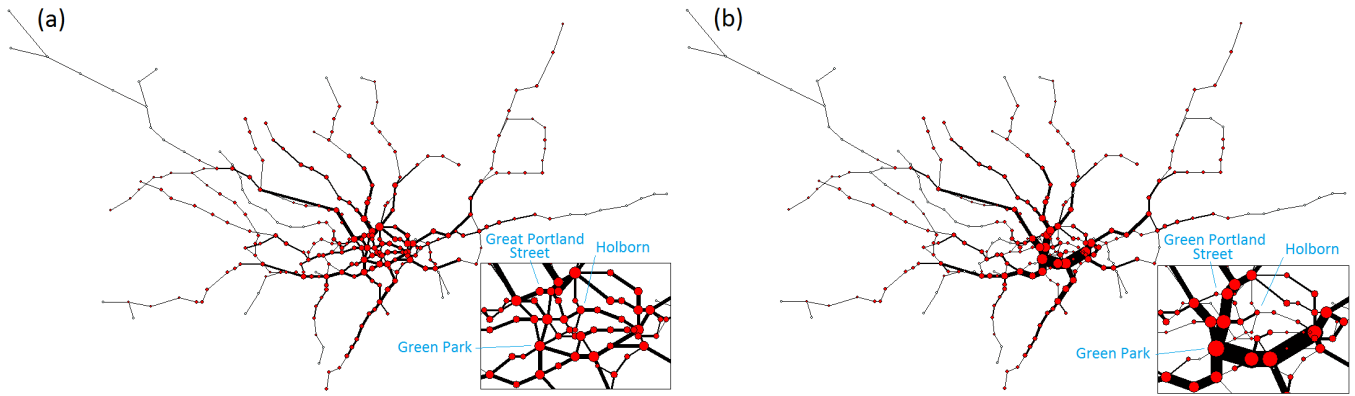, width=\linewidth}}
\end{center}
\caption{
Optimized traffic on the London subway network. A total of 218 real passenger source-destination pairs are optimized, corresponding to 5$\%$ of data recorded by the Oyster card system between 8:30am - 8:31am on one Wednesday in November 2009~\cite{subwayData}. The network consists of 275 stations. The corresponding costs are (a) $\cH\propto\sum_i \flowi^2$, and (b) $\cH\propto\sum_i \flowi^{0.5}$. Red nodes correspond to stations with non-zero traffic. The size of each node and the thickness of each edge are proportional to traffic through them. Insets: zoom on the central region. Nodes are drawn according to their geographic position.
}
\label{fig_subway}
\vspace{-0.4cm}
\end{figure*}

Similar experiments were carried out on the global airport network~\cite{airportData}. Applying the optimization algorithm~(\ref{eq_a}-\ref{eq_b}) to the data one obtains the results presented in~Fig.~2 and Fig.~5. Similar trends to those of the subway network are observed: air-traffic consolidates at airports that are on main routes in the case of $\gamma=0.5$, such as Frankfurt, Toronto and Beijing; while several popular airports such as Tokyo, Newark and Hong Kong show a reduced air-traffic in the case of $\gamma=0.5$ represented by the red line. Table~1 shows the cost obtained by our algorithm when $\gamma=2$ is $56\%$ lower than that obtained by the Dijkstra's shortest paths, with a slight increase in path length of $6.2\%$. This may be due to the availability of a large number of alternative paths in airport network. We note that a lower cost is also achieved in the cases of $\gamma=0.5$.
These results show that our algorithm optimizes a given generic cost, at a price of modest increase in the average path length.

To evaluate the performance of the suggested algorithm (with $\gamma=2$ only) we compared our results against those obtained using a representative state-of-the-art congestion-aware routing algorithm, which we call the \emph{min-cap}~(MC) algorithm~\cite{shahrokhi90}, based on multi-commodity flow. As the latter aims to optimize a linear cost, we have introduced a tunable parameter $\alpha$ such that the quadratic cost is optimized by an extensive search for an optimal $\alpha^*$ (see \emph{SI~Appendix}~Fig.~S8). Details are found in \emph{SI~Appendix}~Sec.~S4. We emphasize that \emph{this comparison is limited to congestion-aware algorithms} ($\gamma \ge 1$) as we have not identified existing efficient optimization algorithms for concave costs that facilitate route consolidation, e.g. the results shown in Figs.~1(b), 3(b) and 5(b).

Table~1 shows a modest gain in cost over the optimized MC results at individual $\alpha^*$ for each run, far less than the gain obtained with respect to Dijkstra's algorithm. Nevertheless, our algorithm provides a lower energy for all $\alpha$ values, unachievable by the MC algorithm even after fine-tuning (see \emph{SI~Appendix}~Fig.~S8). Our algorithm also results in shorter average path length $L$ in addition to lower cost $E$ in random regular graphs (see \emph{SI Appendix}~Table~S1), used as a controlled benchmark problem. Moreover, it is distributed, principled, does not require fine-tuning of free parameters and, most importantly, has the flexibility to accommodate any (non-pathological) cost function designed to address specific needs.

\vspace{-0.4cm}
\subsection{Path Adaptivity}

Figure~4 illustrates the adaptivity of our algorithm after removing the London subway station ``Bank" (black node). Nodes and edges which show an increase (decrease) in optimized traffic are colored red (blue), respectively, with their size and thickness proportional to the magnitude of increase (decrease). Nodes and edges with no traffic changes are in white and black, respectively. In the case of optimization using $\gamma\!=\!2$, the original traffic through ``Bank" is re-routed either via ``Embankment" or  via ``Old Street". This re-distribution of traffic cannot be achieved by ordinary algorithms such as routing tables, shortest-path or minimal weight routing without taking into account the interaction between paths.

On the other hand, in the case of $\gamma=0.5$, almost all the original traffic through ``Bank" is diverted to ``Old Street". As the original traffic via ``Bank" is substantial (see inset of Fig.~3(b)), significant changes at some stations have to be made, although only a small number of stations are subject to re-routing compared to the case of $\gamma=2$.

\begin{figure*}
\begin{center}
\vspace{-0.4cm}
\centerline{\epsfig{figure=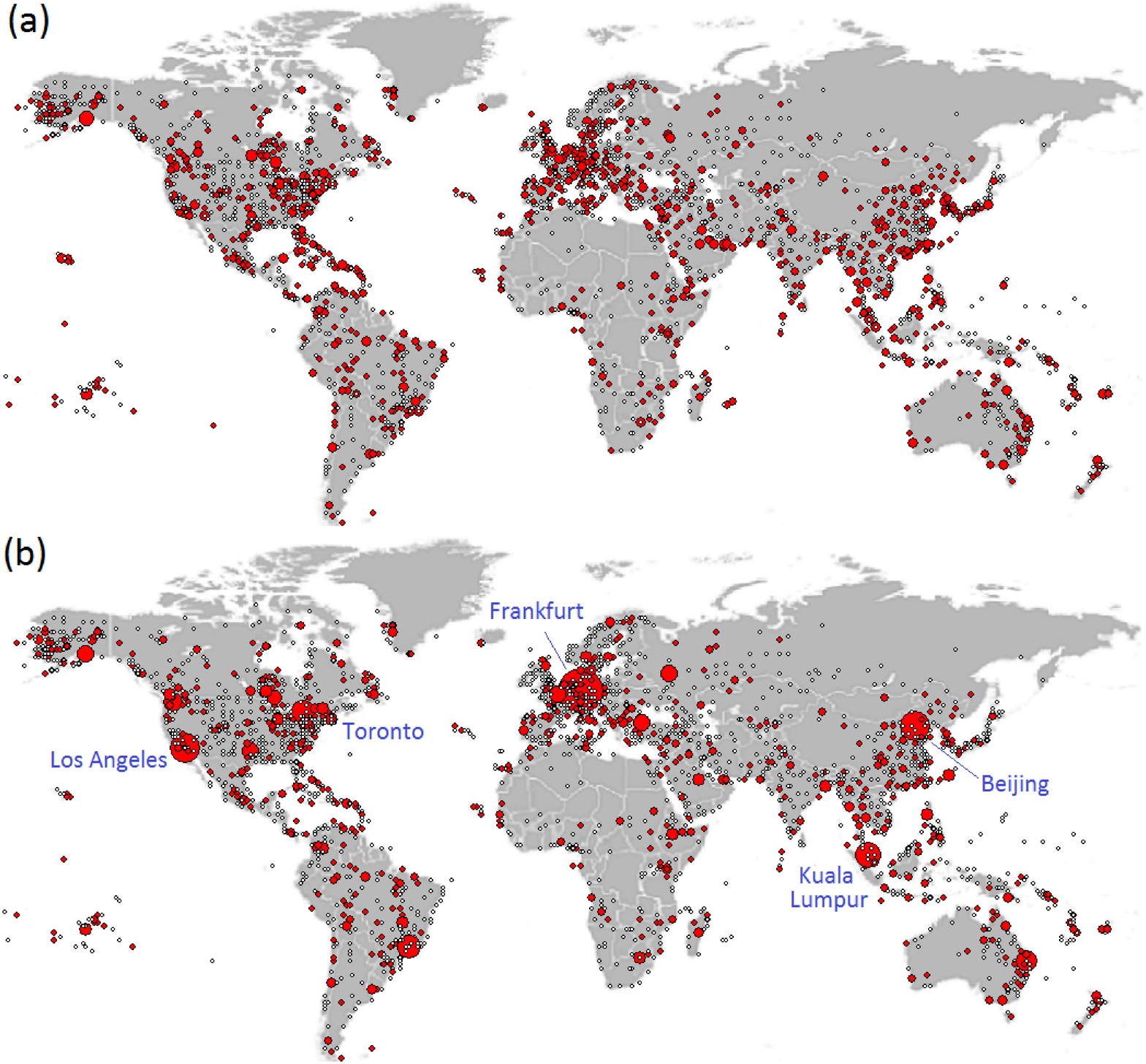, width=0.59\linewidth}}
\end{center}
\vspace{-0.3cm}
\caption{
Optimized traffic at individual airports of the global air network. A total of 2480 airports constitute nodes while edges represent the existence of direct flights between airport pairs~\cite{airportData}. Since the real demand in terms of source-destination pairs is unavailable, it was artificially generated by selecting a set of 300 randomly drawn source-destination pairs. Red nodes correspond to airports with non-zero traffic; the size of nodes indicates the air-traffic through particular airports, edges are omitted for clarity. (a) For $\cH\propto\sum_i\flowi^{2}$ traffic is routed to be almost uniformly distributed to reduce congestion. (b) For $\cH\propto\sum_i\flowi^{0.5}$ air-traffic consolidates at the main hubs.
}
\label{fig_airport}
\end{figure*}

\vspace{-0.4cm}
\subsection{Macroscopic Behavior in Routing}

In addition to the microscopic solutions obtained, we would like to explore the macroscopic behavior of the system. We first examine the dependence of average path length $\avg{L}$ on the number of interacting communications $M$. Random regular networks, Erd\"os-R\'enyi (ER) and scale-free (SF) graphs are studied as they serve as standard benchmark problems and resemble overlay networks on the Internet.
Theoretical results are obtained by solving numerically a set of recursive equations described in \emph{SI~Appendix}~Sec.~S1.4; simulation results are obtained using Eqs.~(\ref{eq_a}) and (\ref{eq_b}).
The inset of~Fig.~6(a) shows  results obtained for random regular graphs. Two remarkable phenomena are observed for both $\gamma=2$ or $\gamma=0.5$: (i) average path length $\avg{L}$ peaks at intermediate $M$ instead of increasing monotonously; (ii) it approaches asymptotically the shortest path $L_1$ as $M\to\infty$ (formally, the value of $\avg{L}$ when $M=1$). Small deviations between theory and simulations are due to finite size effects.

\begin{figure}
\centerline{\epsfig{figure=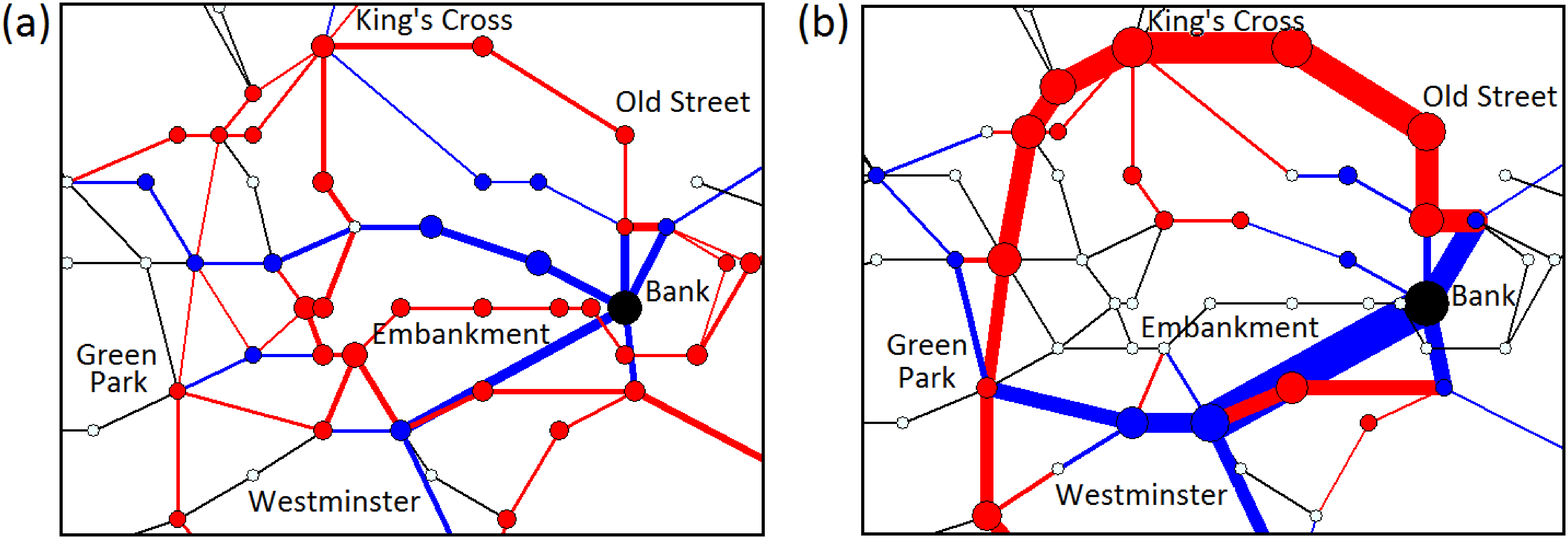, width=\linewidth}}
\caption{
Changes in optimized traffic in the central London subway network after the removal of the station ``Bank" (black node). The corresponding costs are (a) $\gamma=2$ and (b) $\gamma=0.5$. Nodes and edges which show an increase (decrease) in traffic  appear in red (blue), where their size and thickness correspond to the magnitude of increase (decrease). Nodes and edges with no traffic changes appear in white and black, respectively. Passengers source-destination pairs are identical to those of~Fig.~3, except for the removal of pairs starting or ending destinations in ``Bank".
}
\label{fig_delete}
\end{figure}

The observed non-monotonic trends imply the existence of interesting routing phenomena. In the case of $\gamma=2$, it implies that the system is very sensitive to congestion in the intermediate range of $M$. Particularly when $M$ is small, many communications are routed through longer routes as they face stiff competition for shorter ones. However, as $M$ increases further, traffic become more homogeneous and $\avg{L}$ decreases since communications are routed via shorter routes as longer ones are equally congested, matching the experience of frustrated drivers on congested roads. This is reflected in the lower cost obtained by our algorithm in comparison with Dijkstra algorithm, which peaks at 20\% for intermediate $M$ as shown in~Fig.~6(b). Similar trend is observed for $\gamma\!=\!0.5$ as different communications co-operate to share routes in the intermediate range of $M$. As $M$ increases further, traffic becomes more homogeneous and there is less advantage to prefer a busy but longer route, making shorter routes more cost-effective. We note that the peak in $\avg{L}$ for the case of $\gamma=2$ occurs at a smaller $M$ value compared to $\gamma=0.5$, implying that traffic homogeneity is achieved at smaller $M$ in the case of $\gamma=2$.

While similar behaviors are observed for ER graphs (see \emph{SI Appendix} Sec.~S6), SF networks show a much slower decrease of $\avg{L}$ after attaining its maximum, possibly due to the intrinsic node degree inhomogeneity which leads to traffic inhomogeneity even at large $M$. This suggests that shortest-path routing is effective when $M$ is large and topology is homogeneous, but not in networks with high degree variability.

The scaling property of path lengths is shown in~Fig.~6(a). Rescaled path length $(\avg{L}-L_1)(N/\log N)$ with $\gamma=2$ at system sizes $N=100, 200, 500$ and $1000$, plotted as a function of the rescaled number of communication, $M/(N/\log N)$ fall on top of each other almost identically. A similar data collapse is also observed in ER graphs shown in \emph{SI Appendix} Sec.~S4. It implies that the non-monotonic behavior observed for path lengths, and thus the network sensitivity to congestion, depend on $M$ and $N$ only through $M/(N/\log N)$. The latter is roughly proportional to the average traffic on a node since $\log N$ is proportional to the average shortest distance between any two nodes in random regular networks~\cite{newman01,kim04} and ER graphs~\cite{bollobas85}. In other words, the optimal behavior of routing on these graphs depends only on the average node traffic, regardless of system size and number of communications. The rescaling also appears in the reduced cost obtained by our algorithm as shown in~Fig.~6(b). Note that theoretical results have been obtained in the infinite system limit; finite $N$ values have been introduced here merely to determine the scaling properties of $M$.

We have also examined the fraction of idle node as a function of $\gamma$. This revealed a phase transition, an abrupt change in the fraction of idle nodes around the $\gamma=1$ value (see \emph{SI Appendix} Fig.~S11 and Sec.~S7). The implication is that even a small change in the power $\gamma$ is sufficient to effectively power down unnecessary routers or close redundant subway stations, with little impact on the cost or average route length.

\vspace{-0.3cm}
\section{Discussion}

\begin{figure}
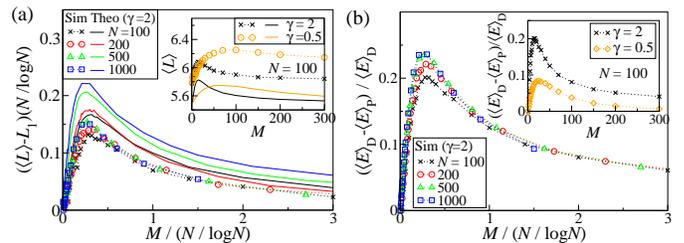

\centerline{\epsfig{figure=pathlength3.eps, width=0.5\linewidth}
\epsfig{figure=energy2.eps, width=0.5\linewidth}}
\caption{
Dependence of the optimized state on the number of communications. (a) The rescaled path length $(\avg{L}-L_{1})(N/\log N)$ and (b) the cost difference $(\avg{E}_D-\avg{E}_P)/\avg{E}_D$ (D and P stand for the Dijkstra algorithm and our algorithm respectively), as a function of the rescaled number of communications $M/(N/\log N)$, for random regular graphs with $N=100, 200, 500, 1000$ and $k=3$; results were obtained for $\cH\propto\sum_i\flowi^{2}$. The value of $L_1$ in (a) corresponds to the value of the shortest path$\avg{L}$. Insets: (a) $\avg{L}$ and (b) $(\avg{E}_D-\avg{E}_P)/\avg{E}_D$ as a function of $M$ for $N=100$ on random regular graphs of degree $k=3$, with cost exponents $\gamma=2$ and $\gamma=0.5$. The error bars for simulation results are of the order of the symbol size. All simulation results are averaged over 2000 realizations.
}
\label{fig_pathLength}
\end{figure}

Optimal routing is one of many hard problems on networks that one should tackle in order to use limited and usually overstretched resources efficiently. The common characteristic of these problems is their global nature and thus the difficulty in solving them at both macroscopic and microscopic levels with limited computational resources. By applying methods from the physics of interacting polymers and disordered systems we obtained typical properties of routing problems and derive a readily applicable, principled, generic, distributed and adaptive routing algorithm. Improvements over state-of-the-art algorithms in the intermediate traffic regime where $M\sim~N\log~N$ are considerable but are modest in the very sparse and dense traffic regimes. These findings will have direct impact on a number of different research areas of practical and societal relevance, from traffic to communication and logistics; but more importantly, 
may open the way for solving many other crucial and non-localized problems on networks.




\vspace{-0.3cm}
\begin{acknowledgments}
This work is supported by EU FET project STAMINA (FP7-265496), Royal Society Exchange Grant IE110151 and Research Grants Council of Hong Kong (605010).
\end{acknowledgments}





\end{article}








\end{document}